\documentclass[paper]{JHEP3} 

\usepackage{epsfig,multicol,bbm}

\newcommand\fverb{\setbox\fverbbox=\hbox\bgroup\verb}
\newcommand\fverbdo{\egroup\medskip\noindent%
			\fbox{\unhbox\fverbbox}\ }
\newcommand\fverbit{\egroup\item[\fbox{\unhbox\fverbbox}]}
\newbox\fverbbox

\newcommand{\be}{\begin{eqnarray}}
\newcommand{\ee}{\end{eqnarray}}

\title{WMAP 7 year constraints on CPT violation from large angle CMB anisotropies}

\author{A. Gruppuso$^{a,b}$, P. Natoli$^{c,a,d}$, N. Mandolesi$^{a}$, A. De Rosa$^{a}$, F. Finelli$^{a,b}$, F. Paci$^{e}$
\\
	$^{a}$ INAF/IASF-BO, Istituto di Astrofisica Spaziale e Fisica Cosmica di Bologna,\\
	via Gobetti 101, I-40129 Bologna, Italy
\\
	$^{b}$ INFN, Sezione di Bologna,
	Via Irnerio 46, I-40126 Bologna, Italy
\\
	$^{c}$ Dipartimento di Fisica and Sezione INFN, Universit\`a degli Studi di Ferrara, \\
	via Saragat 1, I-44100 Ferrara, Italy
\\
	$^{d}$ ASI Science Data Center, c/o ESRIN,
	via G. Galilei, I-00044, Frascati, Italy
\\
	$^{e}$ Laboratoire Astroparticule \& Cosmologie (APC), CNRS, Universit\'e 
	Paris Diderot, 10, rue A. Domon et L. Duquet,
	75205 Paris Cedex 13, France
\\
	E-mail: \email{gruppuso@iasfbo.inaf.it}, \email{paolo.natoli@gmail.com}, \email{mandolesi@iasfbo.inaf.it}, 
	\email{derosa@iasfbo.inaf.it}, \email{finelli@iasfbo.inaf.it},  \email{fpaci@apc.univ-paris7.fr}}

\received{\today} 		
\accepted{\today}		

\abstract{We constrain the rotation angle $\alpha$ of the linear polarization of CMB photons using  the large angular scale (up to $\sim 4^\circ$) signal in WMAP 7 year data. At these scales, the CMB polarization pattern probes mostly the reionization era. A genuine rotation may be interpreted as cosmological birefringence, which is a well known tracer of new physics, through the breakdown of fundamental symmetries. Our analysis provides $\alpha = -1.6^{\circ} \pm 1.7^{\circ} \, (3.4^{\circ})$ at $68\%$ ($95\%$) C.L. for the multipoles range $\Delta \ell = 2-47$ not including an instrumental systematic uncertainty that the WMAP team estimates at $\pm 1.5^\circ$. 
This improves the bound obtained by WMAP team (Komatsu et al., 2010). Moreover we show, for the first time at low multipoles, the angular power spectrum $\alpha_\ell$ in search of a possible scale dependence of the birefringence effect. Our findings are compatible with no detection at all angular scales probed here. We finally forecast the capabilities of Planck in tightening the present constraints.}

\keywords{CMBR theory, CMBR experiments}

\begin{document} 


\section{Introduction}
\label{intro}

The Cosmic Microwave Background (CMB) can be used to constrain fundamental symmetry breaking. Under the Standard Model of particle physics, the CMB temperature to $B$ mode polarization correlation ($TB$) and the $E$ mode to $B$ mode polarization correlation ($EB$) are expected to vanish since parity is conserved \cite{Lue:1998mq,saito}. 
The latter prediction follows from the purely electromagnetic, charge blind nature of the physics of CMB anisotropies \cite{carroll90,carroll91} and becomes observable due to the opposite handedness of $(T,E)$ vs $B$ modes \cite{kks-seljak}. A non null measurement of either $TB$ or $EB$ would thus hint at new physics beyond the Standard Model (or suggest a careful reanalysis of the dataset in search for un-removed systematic effects).

A number of models predict non-standard parity (P) and CP violations (where C stands for charge conjugation), as well as CPT violations (T standing for time reversal) and the associated breakdown of Lorentz invariance 
\cite{greenberg}. Parity violations in the electromagnetic sector are associated with an \textit{in vacuo} rotation of the linear polarization direction of a photon \cite{carroll90,carroll91}, an effect known as cosmological birefringence. 
Several tests have been set forth, either in
earthly and orbital laboratories \cite{bluhm-mewes} or through cosmological observations \cite{carroll90,carroll91,amelino,Alighieri:2010pu}. These violations
should also impact the CMB, whose statistical properties are derived under the assumption of symmetry conservation. The CMB is a powerful probe for such analyses for two main reasons.  
First, it is generated in the early universe, when the physics at the stake was not obviously identical to present. 
Secondly, the long look-back time of CMB photons may render tiny violations to the electromagnetic Lagrangian observable, since such effects usually accumulate during propagation. 

CMB polarization arises at two distinct cosmological times: the recombination epoch ($z \simeq 1100$) and the reionization era at $z \lesssim 11$ \cite{dodelson}. When the CMB field is expanded in spherical harmonics, the first signal mostly shows up at high multipoles, since polarization is generated through a causal process and the Hubble horizon at last scattering only subtends a degree sized angle. The later reionization of the cosmic fluid at lower redshift impacts the low $\ell$ instead. These two regimes need to be taken into account when probing for cosmological birefringence, since they can be ascribed to different epochs and, hence, physical conditions.


Historically, the effect has been first constrained by measuring polarized light from high redshift radio galaxies and quasars \cite{carroll90,carroll91,Alighieri:2010pu,carroll97,Nodland,Eisenstein,Leahy}, see \cite{Alighieri:2010eu}
for a new analysis on ultra-violet polarization of distant radio galaxies.  
For recent, polarization oriented CMB observations \cite{Pryke:2008xp,Komatsu:2010fb,Piacentini:2005yq,Montroy:2005yx} it has become feasible to measure $TB$ and $EB$ correlations, other than $TT$, $TE$ and $EE$ correlations. While no detection has been claimed to date, polarization data have been used to derive constraints on the birefringence angle \cite{Komatsu:2010fb,Feng:2006dp,Cabella:2007br,Wu:2008qb}. 

The goal of this paper is to constrain the cosmological birefringence angle at large angular scale considering the latest WMAP data release.
Our method naturally provides also a scale dependence for such an angle.

In the current analysis we do not address the issue of the systematic effects whose detailed analysis goes beyond the scope of the paper. 
Our new results are based on the WMAP data and we do rely on their original estimate of the uncertainty on the birefringence angle coming from systematic effects \cite{Komatsu:2010fb}. 
For what concerns the Planck forecast we provide, we cannot give an estimate on the errors coming from Planck detectors since those are not yet publicly available.
However, following  \cite{Pagano:2009kj}, we will provide an estimate of the level of accuracy that polarimeter calibration has to satisfy in order to make the error
coming from this systematic effect smaller than the statistical one.

In the limit of constant birefringence angle, $\alpha$, the angular power spectra of CMB anisotropies, assuming $C_{\ell}^{TB} = C_{\ell}^{EB} = 0 $, are given by
 \cite{Lue:1998mq,Feng:2006dp,Feng:2004mq,Komatsu:2008hk} \footnote{See \cite{Galaverni:2009zz,Finelli:2008jv} as an example of computation that takes into account the time dependence of $\alpha$ in a specific model of pseudoscalar fields coupled to photons. See \cite{Li:2008tma,Kamionkowski:2010ss,Gubitosi:2011ue} as examples of non-isotropic birefringence effect.},

\begin{eqnarray}
\langle C_{\ell}^{TE,obs} \rangle &=& \langle C_{\ell}^{TE} \rangle \cos (2 \alpha) \, ,
\label{TEobs} \\
\langle C_{\ell}^{TB,obs} \rangle &=& \langle C_{\ell}^{TE} \rangle \sin (2 \alpha) \, ,
\label{TBobs} \\
\langle C_{\ell}^{EE, obs} \rangle &=& \langle C_{\ell}^{EE} \rangle \cos^2 (2 \alpha) + \langle C_{\ell}^{BB} \rangle \sin^2 (2 \alpha) \, ,
\label{EEobs} \\
\langle C_{\ell}^{BB, obs} \rangle &=& \langle C_{\ell}^{BB} \rangle \cos^2 (2 \alpha) + \langle C_{\ell}^{EE} \rangle \sin^2 (2 \alpha) \, ,
\label{BBobs} \\
\langle C_{\ell}^{EB, obs} \rangle &=& {1 \over 2} \left( \langle C_{\ell}^{EE} \rangle + \langle C_{\ell}^{BB} \rangle \right) \sin (4 \alpha) \, .
\label{EBobs}
\end{eqnarray}
These equations follow from
\begin{eqnarray}
a_{\ell m}^{T,obs} &=& a_{\ell m}^{T} + a_{\ell m}^{n,T} \, , \label{almT} \\
a_{\ell m}^{E,obs} &=& a_{\ell m}^{E} \cos (2 \alpha) +  a_{\ell m}^{B} \sin (2 \alpha) + a_{\ell m}^{n,E} \, , \label{almE}\\
a_{\ell m}^{B,obs} &=& a_{\ell m}^{B} \cos (2 \alpha) +  a_{\ell m}^{E} \sin (2 \alpha) + a_{\ell m}^{n,B} \, , \label{almB}
\end{eqnarray}
where we have explicitly written the contribution due to residual instrumental noise present on the maps. 
As discussed above, the amount of rotation depends on the effective travel time of the photon and the effect is scale (and, hence, harmonic multipole $\ell$) dependent \cite{liu}.  
In the present paper, we focus on the low $\ell$ contribution, with a twofold aim: 
from the one side, we provide a new constraint on $\alpha$ for low $\ell$ using, for the first time (to our knowledge) a linear estimator in combination with an optimal angular power spectrum estimator (APS) for such a regime
\footnote{We stress here that, strictly speaking, it is the adopted APS estimator to be optimal. In particular we have considered the QML method which is called optimal  \cite{Tegmark:1996qt,Tegmark:2001zv} since it is unbiased and minimum variance (it provides the smallest error bars provided by the Fisher-Cramer-Rao inequality). See also Subsection \ref{simulations} and \ref{datset} and reference therein.}; 
from the other side, we give the first measurement of  the spectrum of $\alpha$ at low $\ell$. Note that the technique used by WMAP team does not allow for possible scale dependence of $\alpha$. 

Our constraint reads $\alpha = -1.6^{\circ} \pm 1.7^{\circ} \, (3.4^{\circ})$ at $68\%$ ($95\%$) C.L. for $\Delta \ell = 2-47$.
Considering $\Delta \ell = 2-23$ we obtain $\alpha = -3.0^{\circ +2.6^{\circ}}_{\phantom{a}-2.5^{\circ}}$ at $68\%$ C.L. and 
$\alpha = -3.0^{\circ +6.9^{\circ}}_{\phantom{a}-4.7^{\circ}}$ at $95\%$ C.L.. This is the same multipole range considered by the WMAP 
team in \cite{Komatsu:2010fb} (the only other result available in the literature at these large angular scales) 
where  with a pixel based likelihood analysis they obtain $\alpha^{{\rm WMAP} \, 7 yr} = -3.8^{\circ} \pm 5.2^{\circ}$ at $68\%$ C.L..


The paper is organized as follows: in Section \ref{methodology}
we present the methodology adopted and describe the  simulations performed and the tests proposed;
in Section \ref{results} we show the results obtained (i.e.\ a new and more stringent constraint of $\alpha$ and its spectrum at low $\ell$). 
In Section \ref{conclusions} we draw our main conclusions.
In Appendix \ref{montecarlo} further details about the behavior of the considered estimators are given.

\section{Methodology}
\label{methodology}
\subsection{Estimators}
\label{estimators}

Assuming that the amplitude of the primordial B-mode is zero, we write down the following linear estimators \cite{Wu:2008qb}
\begin{eqnarray}
  D_{\ell}^{TB} &=& C_{\ell}^{TB,obs} \cos (2 \alpha) - C_{\ell}^{TE,obs} \sin (2 \alpha)  \, , \label{DTB} \\
  D_{\ell}^{EB} &=& C_{\ell}^{EB,obs} - {1 \over 2} \left( C_{\ell}^{BB,obs} + C_{\ell}^{EE,obs} \right)  \sin (4 \alpha) \, . \label{DEB} 
\end{eqnarray}
The linearity of the expressions given in Eqs.~(\ref{DTB}) and (\ref{DEB}) guarantees that the estimators are unbiased (even in the presence of instrumental noise)
if the $C_{\ell}^{Y Y^{\prime},obs}$ (where $Y,Y^{\prime}$ are any combination of $T,E,B$) are obtained through an unbiased angular power spectrum estimator 
\footnote{This is not a trivial property. For example, note that the following non-linear estimator 
$$
\sin ( 2 \alpha )=  C_{\ell}^{TB,obs} / \left[ {(C_{\ell}^{TB,obs})}^2+{(C_{\ell}^{TE,obs})}^2\right]^{1/2}
$$
is always biased, even for a noiseless experiment.}. 


The estimators $D_{\ell}^{TB}$ and $D_{\ell}^{EB}$ have zero expectation value as can be readily seen by averaging Eqs.~(\ref{DTB}), (\ref{DEB}) and using Eqs.~(\ref{TEobs}), (\ref{TBobs}),
(\ref{EEobs}), (\ref{BBobs}), (\ref{EBobs}).
Therefore
\begin{eqnarray}
\langle D_{\ell}^{TB} \rangle &=& 0 \, , \label{aveTB} \\
\langle D_{\ell}^{EB} \rangle &=& 0 \, , \label{aveEB}
\end{eqnarray}
for every angle $\alpha$.

In order to find the most probable angle $\alpha$ compatible with the WMAP 7 year low resolution data, we adopt a Bayesian approach and maximize the log-likelihood defined as
\be
-2 \ln {\cal L}^{X}(\alpha) 
=  \sum_{\ell \ell^{\prime}} D^{X, obs}_{\ell}  {M^{X X}_{\ell \ell^{\prime}}}^{-1} D^{X, obs}_{\ell^{\prime}} 
\, ,
\label{loglike}
\ee
where $X=\rm{TB}$ or $\rm{EB}$, $D^{X,obs}$ stands for Eq.~(\ref{DTB}) or (\ref{DEB}) specialized to observed data (i.e. WMAP 7 year low resolution data in the following) and
$M^{X X}_{\ell \ell^{\prime}} = 
\langle D^{X}_{\ell}  D^{X}_{\ell^{\prime}} \rangle$. 
Note that in Eq.~(\ref{loglike}) we have already taken into account Eqs.~(\ref{aveTB}), (\ref{aveEB}).


\subsection{Description of the simulations performed}
\label{simulations}

In this Section we describe the simulations we have performed in order to sample the likelihood defined in Eq.~(\ref{loglike}).

In principle for each fixed angle $\alpha = \bar \alpha$ one has to simulate a large number of primordial maps, rotate them by a quantity $2 \bar \alpha$
and add a random noise realization compatible with the considered data, as given in Eqs.~(\ref{almT}),(\ref{almE}),(\ref{almB}).
Then an angular power spectrum estimator has to be applied to each of the simulated maps to obtain the corresponding set of $C^{Y^{\prime}Y,obs}_{\ell}$, where $Y^{\prime},Y=\rm{T,E,B}$.
Suc estimates are therefore used to build the matrix $M^{X X}_{\ell \ell^{\prime}} $ for the angle $\bar \alpha$. Once $M^{X X}_{\ell \ell^{\prime}} $ is inverted and contracted with $D^{X,obs}$ (of course computed for the 
same angle $\bar \alpha$), from Eq.~(\ref{loglike}) one obtains ${\cal L } (\bar \alpha)$. In order to sample ${\cal L }$, this procedure has to be repeated for every angle
$\alpha$ belonging to the range of interest $[\alpha_{min},\alpha_{max}]$.
 
In practice, since the likelihood given in Eq.~(\ref{loglike}) is by construction a periodic function of $\alpha$ with a period of $90^{\circ}$, one has to perform a 
MonteCarlo (MC) of $N_{sims}$ simulations for each value of $\alpha$ between $-45^{\circ}$ and $45^{\circ}$.
If for example the step ($\Delta \alpha$) of such a sampling is 1 degree, in principle one has to perform
$91$ MC of $N_{sims}$ simulations each. This is not a short task since the computation time of one of those MC with $N_{sims}=10000$ is around $4$ hours on 64 processors using an
optimal estimator at low resolution (see \cite{Gruppuso:2010nd,Paci:2010wp,Gruppuso:2009ab,Tegmark:1996qt,Tegmark:2001zv} for further details on the implementation of the APS estimator we adopt).

In fact, there is a shortcut to such a strategy that allows one to sample the likelihood performing just one MC independently of the choice of the step $\Delta \alpha$.
Feeding Eqs.~(\ref{almT}),(\ref{almE}),(\ref{almB}) into Eq.~(\ref{DTB}) and (\ref{DEB}), after some algebra one obtains
\begin{eqnarray}
D_{\ell}^{TB} &=&   ( C_{\ell}^{T, nB} + \delta C_{\ell}^{nT, nB})  \cos (2 \alpha) -  ( C_{\ell}^{T, nE} + \delta C_{\ell}^{nT, nE}) \sin (2 \alpha)  \, , \label{DTBnew}   \\
D_{\ell}^{EB} &=& C_{\ell}^{E,nB} \cos^2 (2 \alpha) + \delta C_{\ell}^{nE,nB} -{1 \over 2} \left( C_{\ell}^{E,nE} + \delta C_{\ell}^{nE,nE} + \delta C_{\ell}^{nB,nB} \right)  \sin (4 \alpha)  \, , \label{DEBnew} 
\end{eqnarray}
where $C^{Y^{\prime},nY}_{\ell}$ is the angular power spectrum of the correlation between $Y^{\prime}$ and the noise of $Y$
and $\delta C^{nY^{\prime},nY}_{\ell}$ is the angular power spectrum of the residual noise leftover after the noise bias removal,
i.e. $\delta C^{nY^{\prime},nY}_{\ell} =  C^{nY^{\prime},nY}_{\ell} - b^{nY^{\prime},nY}_{\ell} $ with $b^{nY^{\prime},nY}_{\ell} $ being
the noise bias estimate.
Eqs.(\ref{DTBnew}),(\ref{DEBnew}) indicate clearly that the matrix $M^{X X}_{\ell \ell^{\prime}} $ can be built starting from the random extractions of $C^{Y^{\prime},nY}_{\ell}$ and $\delta C^{nY^{\prime},nY}_{\ell}$
which do not depend on $\alpha$.

\subsection{Power spectra estimates from WMAP 7 data}
\label{datset}

In this Section we describe the data set that we have considered. 
We use the temperature ILC map smoothed at $9.8$ degrees and reconstructed at HealPix\footnote{http://healpix.jpl.nasa.gov/}
 \cite{gorski} resolution\footnote{For the reader who is not familiar with HEALPix convention, $N_{side}=16$
corresponds to maps of $3072$ pixels on the whole sky.} $N_{side}=16$, 
the foreground cleaned low resolution maps and the noise covariance matrix in $(Q,U)$ publicly available at the LAMBDA website
\footnote{http://lambda.gsfc.nasa.gov/} for the frequency channels Ka, Q and V as considered by Larson et al. in \cite{Larson:2010gs} 
for the low $\ell$ analysis of the WMAP data.
These frequency channels have been co-added as follows \cite{Jarosik:2006ib}
\be
m_{tot} = C_{tot} (C_{Ka}^{-1} m_{Ka} + C_Q^{-1} m_Q+ C_V^{-1} m_V)
\, ,
\ee
where $m_{i}$, $C_i$ are the polarization maps  and covariances (for $i=Ka, Q$ and $V$) and
\be
C_{tot}^{-1} = C_{Ka}^{-1} + C_{Q}^{-1} +C_{V}^{-1} 
\, .
\ee
This polarization data set has been extended to temperature considering the ILC map.
We have added to the temperature map a random noise realization with variance of $1 \mu K^2$ as suggested in \cite{Dunkley:2008ie}.
Consistently, the noise covariance matrix for TT is taken to be diagonal with variance equal to $1 \mu K^2$.




We have performed Monte-Carlo simulations in order to properly sample the likelihood given in Eq.~(\ref{loglike}).
A set of $10000$ CMB plus noise sky realizations has been generated: the signal extracted from the WMAP 7 years best fit model, the noise through a Cholesky decomposition of the noise
covariance matrix. 
We have then computed the APS for each simulation by means of our implementation of the QML estimator, namely {\it BolPol} \cite{Gruppuso:2009ab}, keeping record not only 
of the estimated $C_{\ell}^{Y^{\prime}Y}$ but also of signal and noise only extractions. By saving the latter, one may obtain $C^{Y^{\prime},nY}_{\ell}$ and $\delta C^{nY^{\prime},nY}_{\ell}$ given in Eqs.(\ref{DTBnew}),(\ref{DEBnew}).

We considered two different Galactic masks: one for temperature and one for linear polarization, as suggested by the WMAP team \cite{Larson:2010gs}. 
Best-fit monopole and dipole have been subtracted from the observed ILC map through the HealPix routine {\sc remove-dipole} 
\cite{gorski}. 

\section{Results from WMAP 7 year} 
\label{results}

\subsection{Likelihood for $\alpha$}
\label{likelihood}

\begin{figure*}
\includegraphics[width=4.9cm]{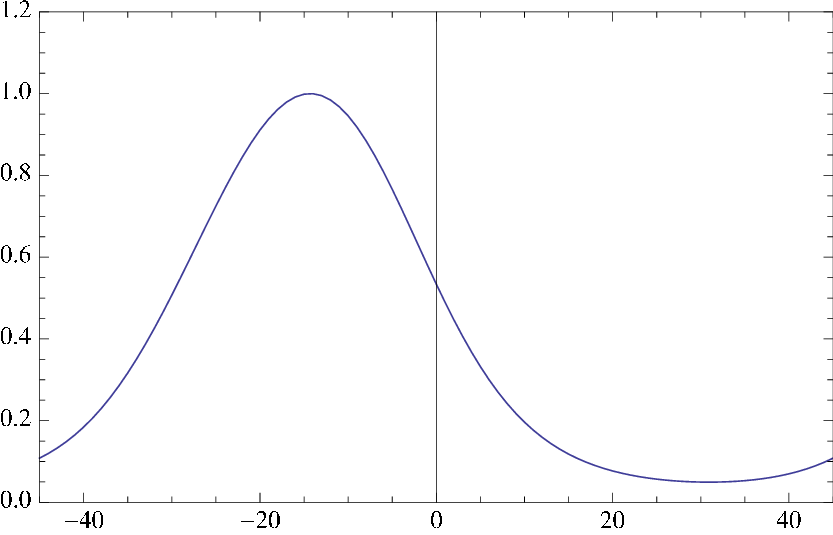}
\includegraphics[width=4.9cm]{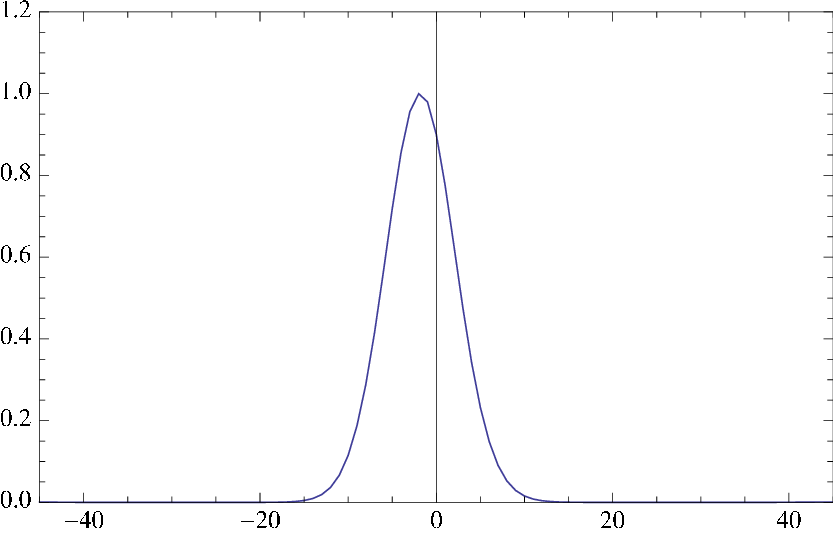}
\includegraphics[width=4.9cm]{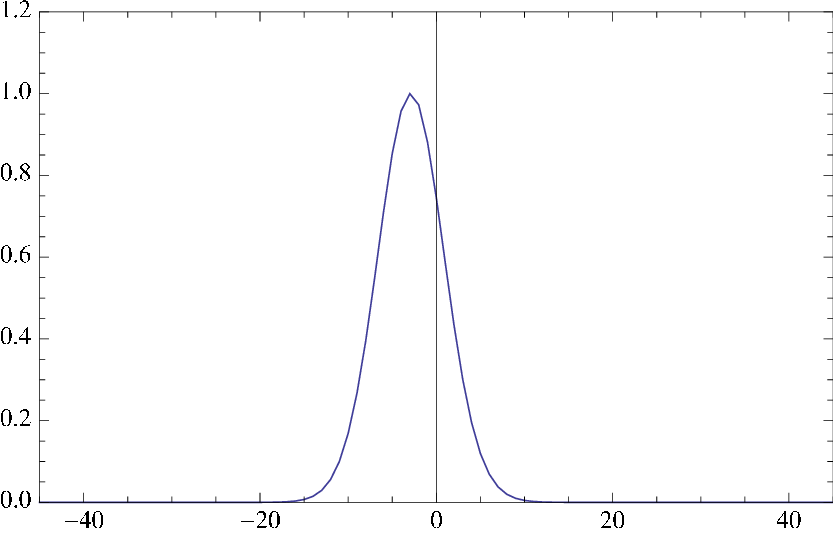}

\includegraphics[width=4.9cm]{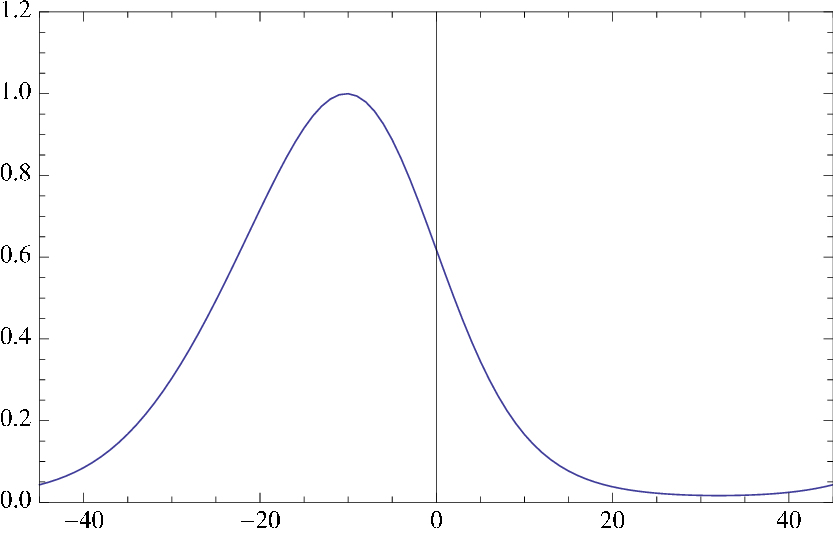}
\includegraphics[width=4.9cm]{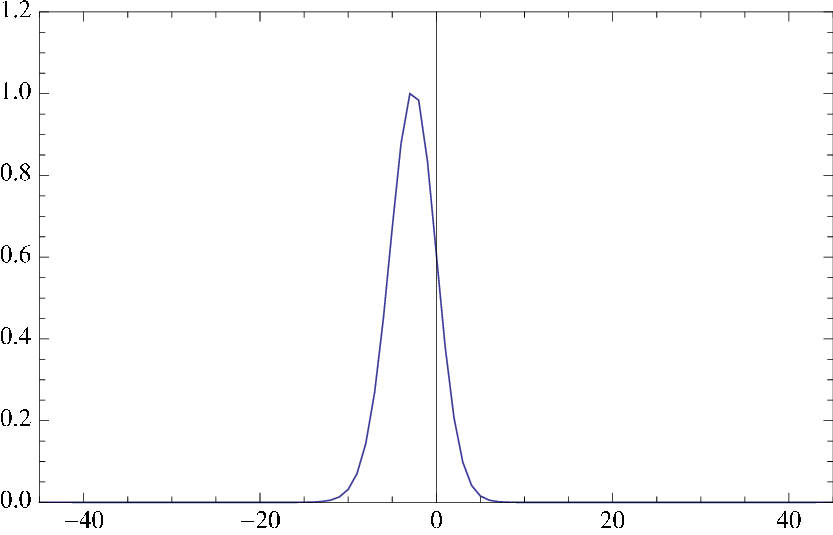}
\includegraphics[width=4.9cm]{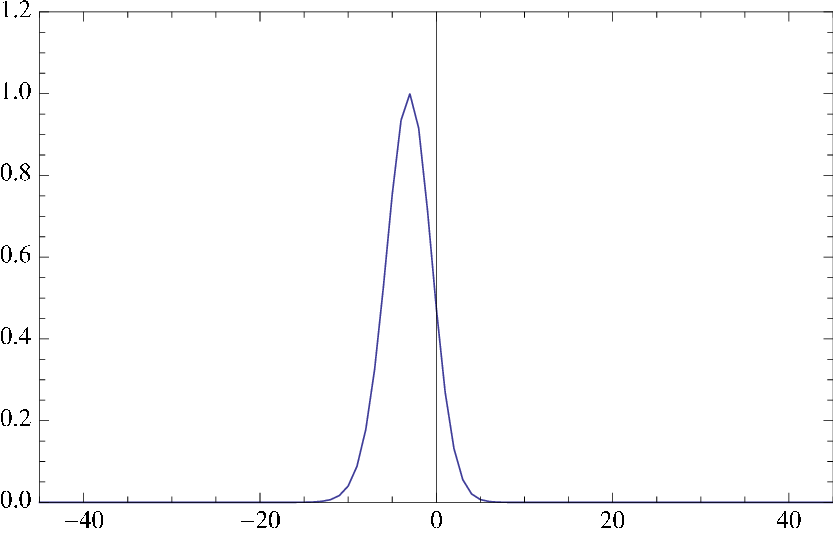}

\includegraphics[width=4.9cm]{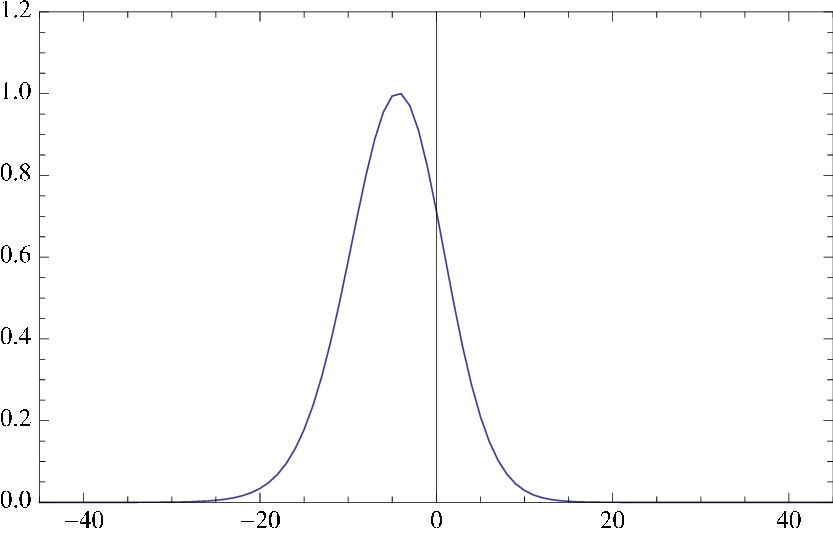}
\includegraphics[width=4.9cm]{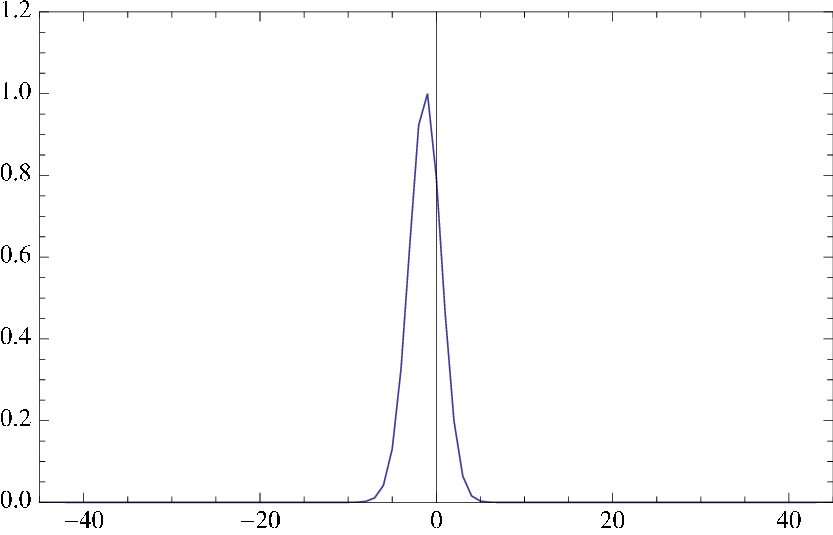}
\includegraphics[width=4.9cm]{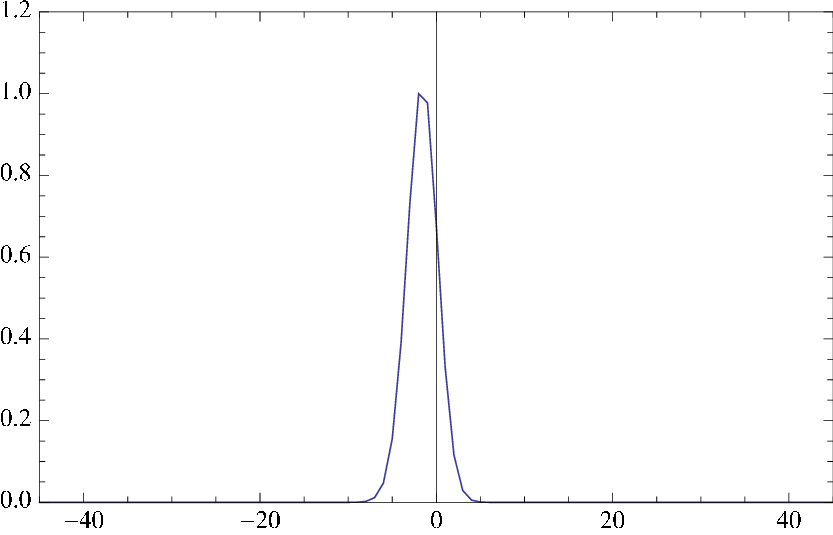}

\caption{Left column of panels: likelihood for $D^{TB}$.
Middle column of panels: likelihood for $D^{EB}$.
Right column of panels: joint likelihood for $D^{TB}$ and $D^{EB}$.
First row $\Delta \ell = 2-6$.
Second row $\Delta \ell = 2-23$.
Third row $\Delta \ell = 2-47$.
Each panel represents the likelihood (normalized to unity at the peak) vs $\alpha$, expressed in deg.
See also the text.} 
\label{likelihoodplot}
\end{figure*}

In Fig.~\ref{likelihoodplot} we show the likelihood distribution we obtain by sampling Eq.~(\ref{loglike}) for X=TB (left column), X=EB (middle column).
Right column of Fig.~\ref{likelihoodplot} shows the joint likelihood of the two estimators computed as
\be
-2 \ln {\cal L}^{joint}(\alpha) = -2  \ln {\cal L}^{TB}(\alpha) -2  \ln {\cal L}^{EB}(\alpha)
\, .
\label{jointloglike}
\ee

The best constraint we obtain for the birefringence angle comes from the joint likelihood for the largest range of multipoles (i.e. $\Delta \ell = 2-47$ corresponding to the 
lower-right panel of Fig.~\ref{likelihoodplot}) and it is given by $\alpha = -1.6^{\circ} \pm 1.7^{\circ} \, (3.4^{\circ})$ at $68\%$ ($95\%$) C.L..
Considering $\Delta \ell = 2-23$ we obtain
$\alpha = -3.0^{\circ +2.6^{\circ}}_{\phantom{a}-2.5^{\circ}}$ at $68\%$ C.L. and $\alpha = -3.0^{\circ +6.9^{\circ}}_{\phantom{a}-4.7^{\circ}}$ at $95\%$ C.L..
This is the same multipole range considered by the WMAP team in \cite{Komatsu:2010fb} where  with a pixel based likelihood analysis 
they obtain $\alpha^{{\rm WMAP} \, 7 yr} = -3.8^{\circ} \pm 5.2^{\circ}$ at $68\%$ C.L..

We note that such results are compatible well within $1 \sigma$ and moreover we argue that the linear estimators 
of Eqs.~(\ref{DTB}) and (\ref{DEB}) provide tighter constraints (at least when coupled with an optimal estimators for APS)
with respect to the technique adopted in  \cite{Komatsu:2010fb}. 

All errors quoted above and in what follows are purely statistical. We notice, incidentally, that the WMAP team quotes an instrumental systematic error on the polarization angle at $\pm 1.5^\circ$ 
of the nominal orientation \cite{Komatsu:2010fb,Page:2003bc,Page:2006hz}. This further uncertainty contributes systematically to our final estimates.

\subsection{Angular spectrum for $\alpha$} 
\label{spectrum}

\begin{figure}
\includegraphics{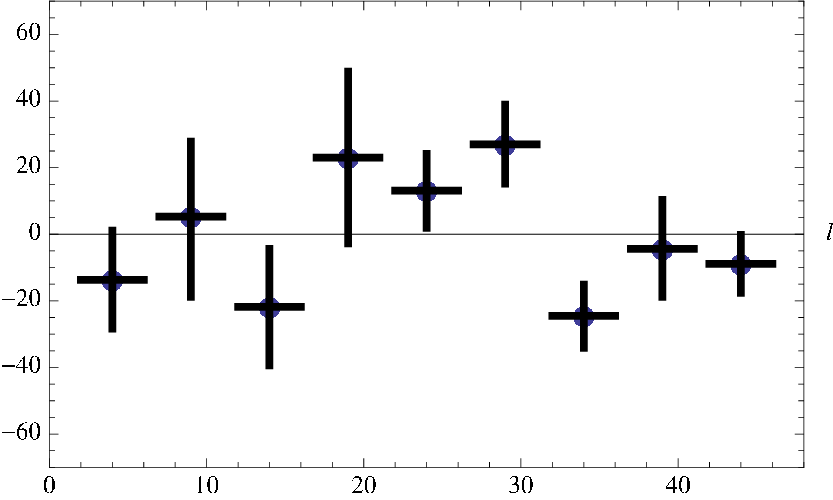}

\caption{Spectrum of $\alpha$ vs $\ell$ obtained from $D^{TB}_{\ell}$. 
The dot represents the peak position and the error bars provides the $1 \sigma$ C.L.. See also the text.} 
\label{spectrumplotTB}
\end{figure}

As performed in \cite{Wu:2008qb} we specialize Eq.~(\ref{loglike}) to a given multipole range $\left[ \ell_{\rm min}, \ell_{\rm max} \right]$,
\be
-2 \ln {\cal L}^{X}_{(\ell_{\rm min},\ell_{\rm max})}(\alpha)
=  \sum_{\ell=\ell_{\rm min}}^{\ell_{\rm max}} \sum _{\ell^{\prime}} D^{X, obs}_{\ell}  {M^{X X}_{\ell \ell^{\prime}}}^{-1} D^{X, obs}_{\ell^{\prime}} 
\, ,
\label{loglikenew}
\ee
where we have made explicit the dependence of $\ell_{\rm min},\ell_{\rm max}$ with only $\ell^{\prime}$ being summed over the entire available multipole range (i.e. from $2$ to $47$).
This allows one to compute the spectrum of $\alpha$, i.e. $\alpha$ vs $\ell$. 
In Fig.~\ref{spectrumplotTB} we show the spectrum we obtain when we use Eq.~(\ref{loglikenew}) for $X=TB$.
The error bars are obtained taking the $1 \sigma$ C.L..

Note that for $X=EB$ at $\ell \gtrsim 8-9$  pure noise fluctuations overtake the signal-noise ones. 
Therefore, for such range of multipoles the sampled likelihoods might not be well featured.
This is not the case for $X=TB$ where this transition happens for $\ell \gtrsim 40$ 
(see Appendix \ref{montecarlo} for further details).
This is the reason why we show the spectrum for $TB$ in Fig.~\ref{spectrumplotTB} but not the analogous for $EB$.
The data for $EB$ are taken into account only in combination with $TB$ when building the joint likelihood, see Fig. \ref{likelihoodplot}.
However, when the lowest multipoles are included in the analysis any possible noisy feature on the ${\cal L}^{EB}$ 
due to higher multipoles is subdominant, as shown by the middle column of Fig.~\ref{likelihoodplot}.

\section{Planck forecasts}
\label{planck}
In this Section we take into account the white noise level for $143$ GHz channel of the Planck mission \cite{blubook} launched into space on the 14th of May of 2009.
As in \cite{Gruppuso:2010nd,Paci:2010wp}, we consider the nominal sensitivity of the Planck $143$ GHz channel, taken as representative of the results which can be obtained after the foreground cleaning from various frequency channels. 
The $143$ GHz channel has an angular resolution of $ 7.1^{\prime}$ (FWHM) and an average sensitivity of $6 \, \mu K \, \, (11.4 \, \mu K) $ per pixel - a square whose side is the FWHM size of the beam - in temperature (polarization), 
after $2$ full sky surveys. 
We assume uniform uncorrelated instrumental noise and we build the corresponding diagonal covariance matrix for temperature and polarization, from which, through Cholesky decomposition we are able to extract noise realizations.
For this low noise level we apply the same procedure adopted for the Monte-Carlo simulations in Subsection \ref{datset}.

\begin{figure}
\includegraphics[width=4.9cm]{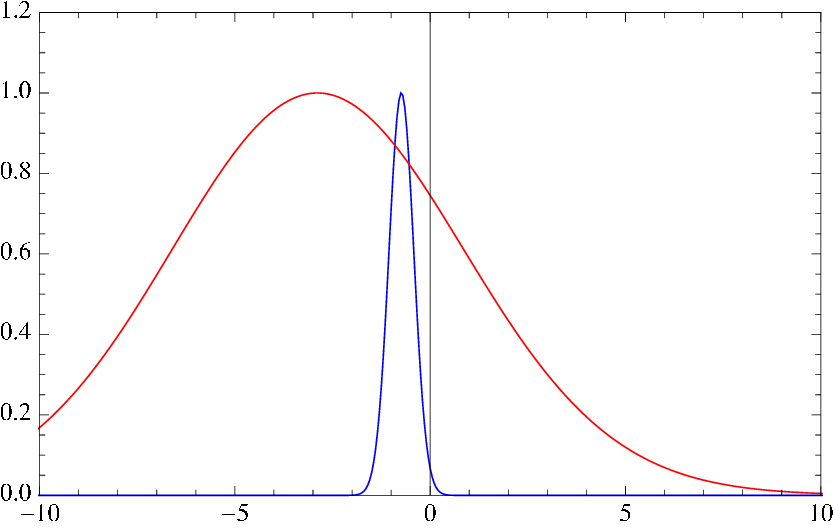}
\includegraphics[width=4.9cm]{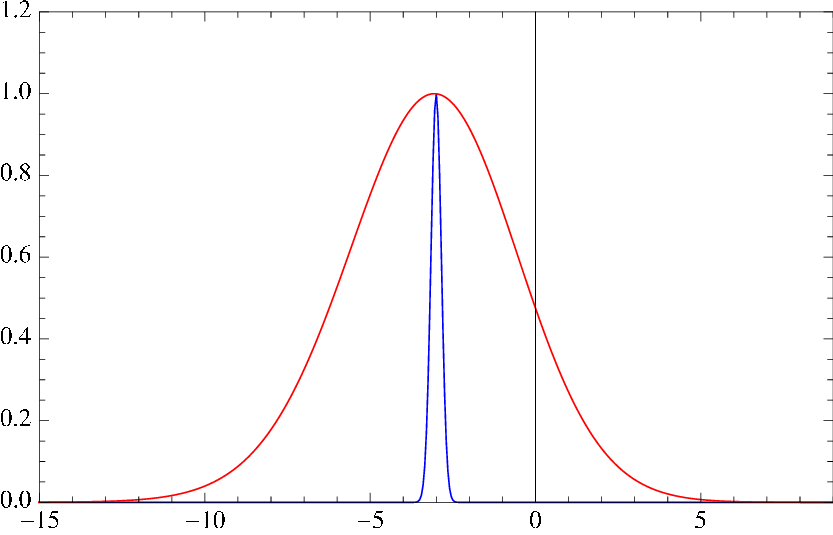}
\includegraphics[width=4.9cm]{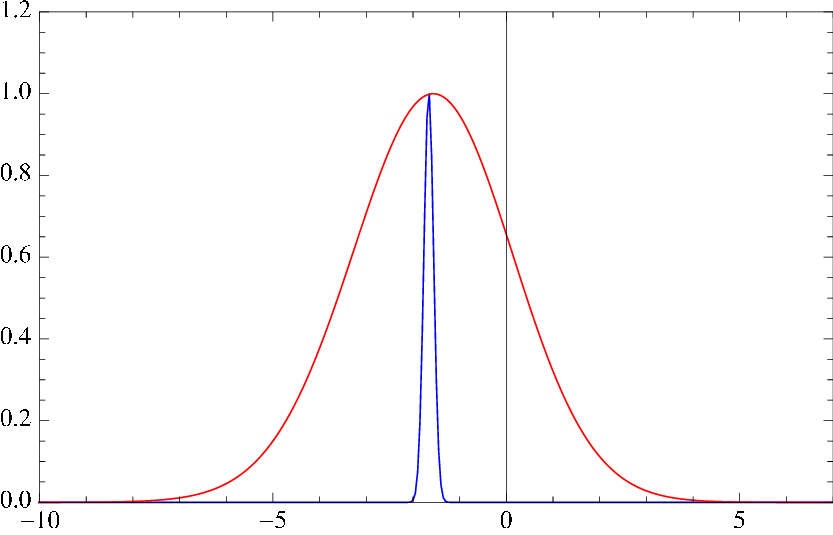}

\caption{Joint likelihood of $\alpha$ for $\Delta \ell= 2 - 16$ in the left panel,  $\Delta \ell= 2 - 23$ in the middle panel
and $\Delta \ell= 2 - 47$ in the right panel. 
Red is for the WMAP data and blue for the Planck forecast. See also the text.} 
\label{comparison}
\end{figure}

In Fig.~\ref{comparison} we show the Planck forecast for the joint likelihood (plotted in blue). In the same panels
we give also the WMAP joint likelihood for comparison (plotted in red). Left panel of  Fig.~\ref{comparison}
is for $\Delta \ell= 2 - 6$, middle panel for $\Delta \ell= 2 - 23$, and right panel for $\Delta \ell= 2 - 47$. The Planck likelihoods are computed as if Planck would have observed the same data as WMAP but with its own noise amplitude.
We estimate an improvement by a factor $~16.3$ on the standard deviations.
Of course the improvements shown in Fig.~\ref{comparison}  are based only on the statistical noise and do not consider the systematic errors that are not publicly available now.
However, following the analysis performed in \cite{Pagano:2009kj} we can infer that the accuracy with which the angles and calibration of the
Planck detectors must be known for the stated Planck forecasts to be dominant in the error budget, is $\Delta \alpha^{syst} \lesssim 0.1^{\circ}$.

\section{Conclusions} 
\label{conclusions}
We have performed an analysis of the WMAP 7 low resolution data in search for a birefringence effect in the CMB. The large angular scale considered here probes the reionization era, as opposed to other higher resolution analyses that are more sensitive to last scattering epoch. The combination of an optimal estimator for the CMB angular power spectra and a linear estimator, Eq. (\ref{DTB}) and (\ref{DEB}), provides constraints in the range $\Delta \ell= 2 - 23$ which are tighter by a factor $\sim 2$ than WMAP's  own analysis, the only other results available in the literature at these large angular scales. 
We report our best constraint $\alpha = -1.6^{\circ} \pm 1.7^{\circ} \, (3.4^{\circ})$ at $68\%$ ($95\%$) C.L. considering $\Delta \ell= 2 - 47$. We also provide, for the first time here, an angular spectrum of the birefringence angle $\alpha$ over the multipole range $\ell \lesssim 50$, detecting no significant deviation from the hypothesis of a scale independent $\alpha$. We have, finally, forecasted the improvement expected from Planck data, finding that it should outnumber WMAP by a factor above $16$ in terms of standard deviations.

\acknowledgments
We acknowledge the use of computing facilities at NERSC. 
We acknowledge the use of the Legacy Archive for Microwave Background Data Analysis (LAMBDA). Support for LAMBDA is provided by the NASA Office of Space Science.
Some of the results in this paper have been derived using the HEALPix \cite{gorski} package.
This work has been done in the framework of the {\it Planck} LFI activities. We acknowledge support by ASI through ASI/INAF agreement I/072/09/0 for the {\it Planck} LFI activity of Phase E2.

\appendix

\section{Montecarlo simulations}
\label{montecarlo}

\begin{figure}

\includegraphics[width=15cm]{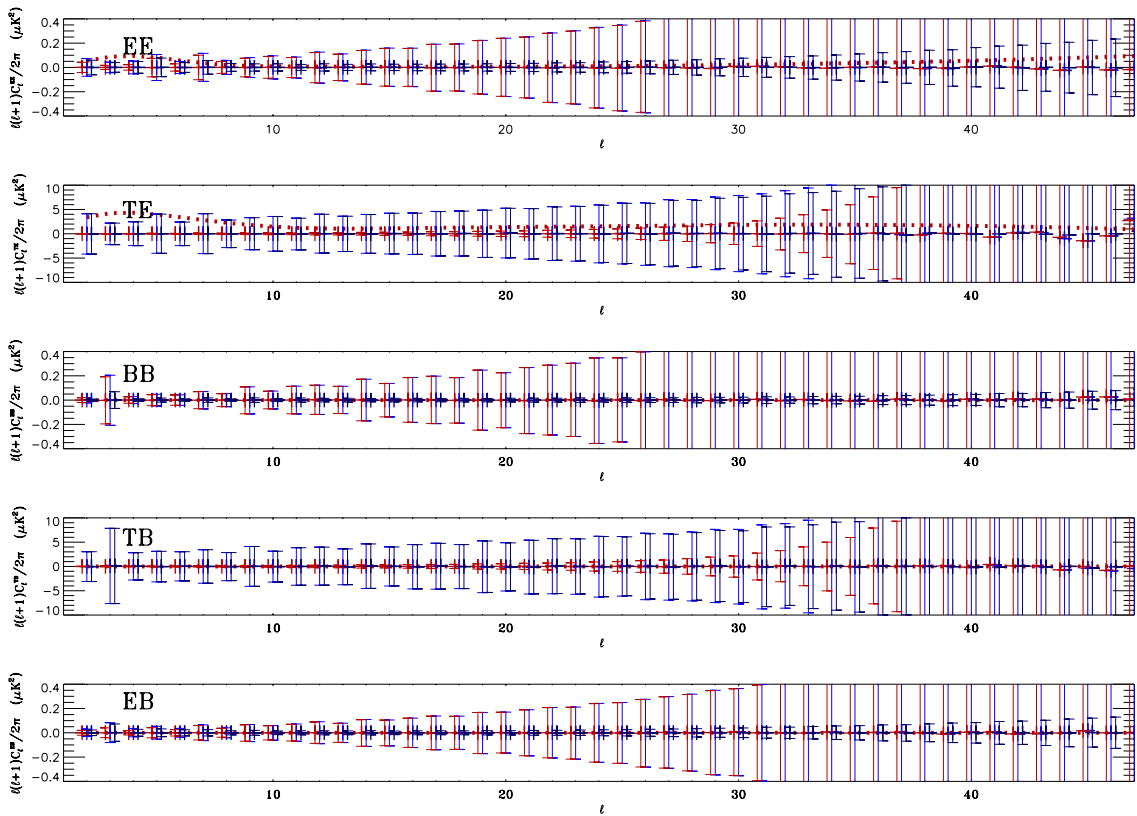}

\caption{Montecarlo for the WMAP 7 yr low resolution data. $\delta C^{nY^{\prime},nY}_{\ell}$ (left vertical bars),
$C^{Y^{\prime},nY}_{\ell}$ (right vertical bars) and their sum (middle vertical bars). 
Dotted lines are for the fiducial CMB power spectrum. 
See also the text.} 
\label{MCWMAP}
\end{figure}

In this appendix we give some details about the angular dependence of the fluctuations entering the definition of our estimators, i.e.  Eqs.~(\ref{DTB}) and (\ref{DEB}).
In Fig.~\ref{MCWMAP} we present the averages and the standard deviation of the MC for the WMAP 7 year low resolution data.
In particular, for each multipole and for each spectrum, we provide information about $\delta C^{nY^{\prime},nY}_{\ell}$ (left vertical bars),
$C^{Y^{\prime},nY}_{\ell}$ (right vertical bars) and their sum (middle vertical bars). 
The averages are zero because the pure signal contribution does not enter the definition of the estimators and has been removed.
It is clear that we have two kinds of behavior in Fig.~\ref{MCWMAP}: for $X=TB$ (that means considering the panels for $TB$ and $TE$) we have
the signal-noise APS above the noise-noise APS up to $\ell \sim 40$ (for the WMAP case) whereas for $X=EB$ (that means considering the panels for $EE$, $BB$ and $EB$)
the signal-noise APS are of the same order as the noise-noise APS up to $\ell \sim 10$ (except for $BB$ whose signal is zero by construction). 
Beyond this scale the noise-noise APS is dominant.

\begin{figure}

\includegraphics[width=15cm]{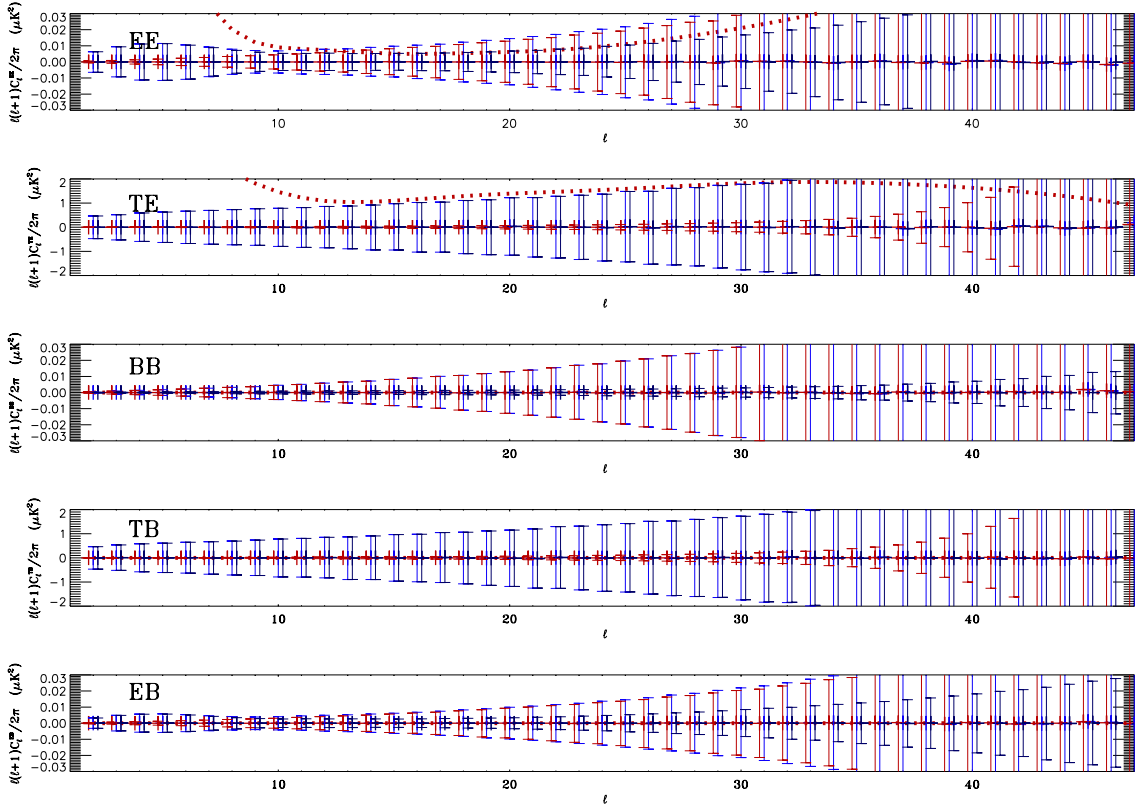}

\caption{Montecarlo for the Planck low resolution data. Same conventions as in Fig.~\ref{MCWMAP}.} 
\label{MCPlanck}
\end{figure}

For the Planck case, see Fig.~\ref{MCPlanck}, the behavior is analogous to what already described for the WMAP case with the following differences:
for $X=EB$ the signal-to-noise ratio is improved for the lowest multipoles but the transition to noise dominated regime happens almost at the same critical multipole,
i.e. $\ell \sim 10-11$ (since the primordial signal vanishes at that scale),
and for $X=TB$ the signal-noise APS are dominant for all the range of multipoles that are accessible at the considered resolution.
Therefore, for the Planck data, when considering $X=TB$ it might be worth to perform this analysis at higher resolution, e.g. $N_{side}=32$.

\end{document}